\documentclass{svproc}
\usepackage{url}
\usepackage{lipsum}
\usepackage{graphicx}
\usepackage{fancyhdr}
\usepackage{amsmath}

\begin{document}
\mainmatter              % start of a contribution
\title{Generalized degeneracies and their resolution in
neutrino oscillation experiments}
\titlerunning{Neutrino oscillations}  % abbreviated title (for running head)
\author{Newton Nath\inst{1,2} \and Srubabati Goswami\inst{1}
\and K.N. Deepthi\inst{1}}
\authorrunning{Newton Nath et al.} % abbreviated author list (for running head)
%Craig Chambers, Kim B. Bruce, and Elisa Bertino}
%
\institute{$^1$Physical Research Laboratory, Navrangpura, Ahmedabad--380 009,  India \\
$^2$Indian Institute  of  Technology,  Gandhinagar,  Ahmedabad--382 424,  India
\email{newton@prl.res.in}
%\texttt{http://users/\homedir iekeland/web/welcome.html}
%\and
%Universit\'{e} de Paris-Sud,
%Laboratoire d'Analyse Num\'{e}rique, B\^{a}timent 425,\\
%F-91405 Orsay Cedex, France
}

\maketitle              % typeset the title of the contribution
\vspace{-7mm}
\begin{abstract}
We  discuss a comprehensive way to study the parameter degeneracies in the form of a generalized degeneracy in the neutrino oscillation experiments. First we describe the various degeneracies 
%present both in probability as well as in event level 
by considering only neutrino run of the long baseline experiment (LBL), NO$ \nu $A. Then we discuss the role of antineutrinos. Later, we present the combined role of T2K (LBL experiment) and ICAL@INO (atmospheric experiment) to resolve these degeneracies. We also discuss the affect of  new physics like non-standards interactions (NSI) on the determination of neutrino mass hierarchy in DUNE.
%
%The abstract should summarize the contents of the paper
%using at least 70 and at most 150 words. It will be set in 9-point
%font size and be inset 1.0 cm from the right and left margins.
%There will be two blank lines before and after the Abstract. \dots
% We would like to encourage you to list your keywords within
% the abstract section using the \keywords{...} command.
\end{abstract}
\section{Introduction}
Standard three-flavor neutrino oscillation paradigm consists of six oscillation parameters, these are ; (i) 3-mixing angles ($\theta_{ij}, j>i=1,2,3$)
, (ii) 2-mass squared differences ($\Delta m^2_{i1}, i=2,3$)  and (iii) the Dirac CP phase $\delta_{CP}$. Almost two decades of neutrino oscillation experiments have measured or given  hints about these parameters. 
%The CP phase $\delta_{CP}$ remains one of the least known parameters among all these. 
Currently, the major three unknowns  in neutrino oscillation physics are, (i) neutrino mass hierarchy, i.e. the sign of $|\Delta m^2_{31}|$($\Delta m^2_{31} > 0$ is known as the normal hierarchy (NH) or $\Delta m^2_{31} < 0$ is  known as the inverted hierarchy (IH)),  (ii) the octant of $\theta_{23}$ ( $\theta_{23} < \pi/4$ is known as the lower octant (LO) or $\theta_{23} > \pi/4$ is  known as the higher octant (HO)) and (iii) the CP phase $\delta_{CP}$ , recently T2K results  hint towards the maximal $\delta_{CP}$ value \cite{Abe:2013hdq}.  
The  (LBL) oscillation experiments like, T2K  \cite{Abe:2015awa}  and NO$ \nu $A \cite{Adamson:2016tbq} which are currently taking data, can provide information on these unknowns. The major obstacles which these LBL experiments have to overcome are the issues of parameter degeneracies i.e. at least two different sets of parameters giving rise to the same oscillation probability. 

In this work, we show in a comprehensive way the parameter degeneracies in the test ($ \theta_{23}\times\delta_{CP} $)-plane for a given set of representative true values for both the hierarchies. Depending on right (R) or wrong (W) values of (hierarchy$-$octant$-\delta_{CP} $), there can be 8-possible solutions.  We show all the possible observed degeneracies, by considering NO$ \nu $A neutrino runs and then we discuss the role of antineutrinos ($ \overline{\nu} $ s) to resolve these degeneracies. We then demonstrate how the addition of T2K and ICAL@INO can help in further  constraining the degenerate solutions.
% and to improve the precision of the parameters. 
Sub-leading effects originating from new physics beyond Standard Model may affect the determination of various unknowns in neutrino oscillation physics. In the near future, this can be probed in the  neutrino oscillation experiments. In this respect, we also present a possible  new physics scenario, namely NSI and discuss its effect on the determination of neutrino mass hierarchy. The oscillation probabilities which are relevant in our study are considered from \cite{Akhmedov:2004ny}. The simulation details and experimental specifications that we considered are given in Ref. \cite{Ghosh:2015ena,Deepthi:2016erc} and the references there in.  The current best fit values and $3\sigma$ ranges,  that we considered in our study  are consistent with  \cite{Gonzalez-Garcia:2014bfa,global_valle}. 
%$$$$$$$$$$$$$$$$$$$$$$$$$$$$$$$$$$$$$
\vspace{-4mm}
\section{Results}
\begin{figure}
\vspace{-0.5cm}
        \begin{tabular}{lr}
               \hspace*{-0.6in}
\includegraphics[scale=0.7]{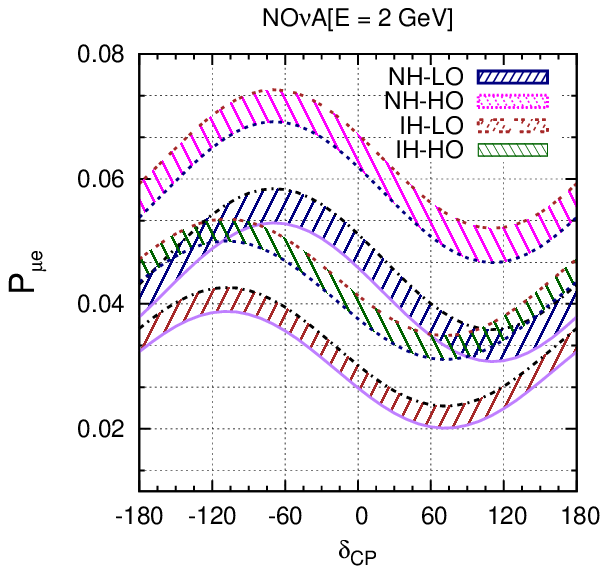}
\hspace*{-0.8in}
\includegraphics[scale=0.7]{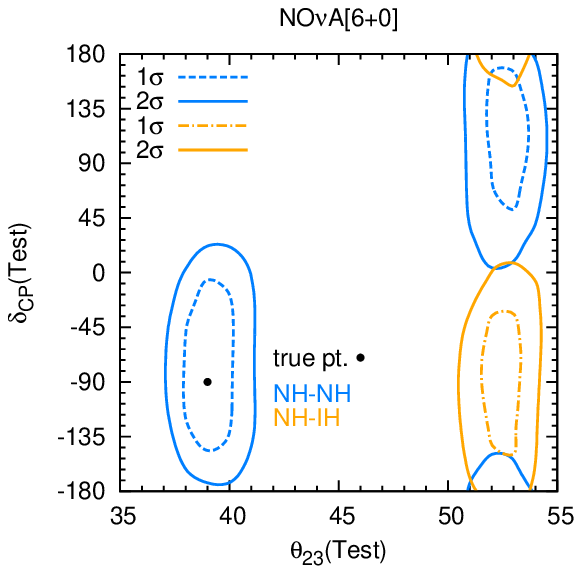}
\hspace*{-1.0in}
\includegraphics[scale=0.7]{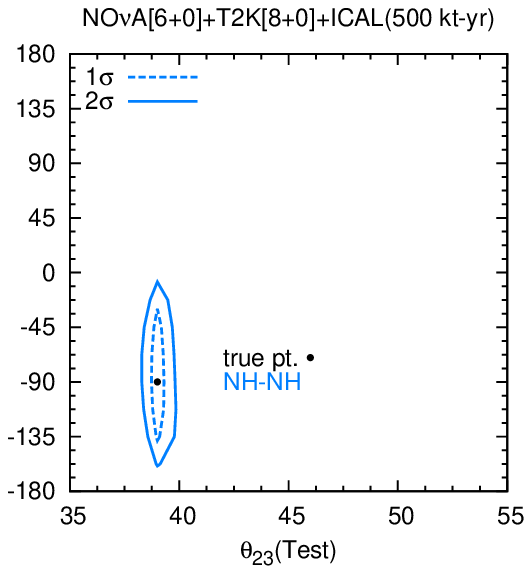}
        \end{tabular}
\vspace{-0.55cm}
\caption{\footnotesize Here, first (second) column shows the degeneracies in probability (event) level.  Whereas third column shows the removal of degeneracies and the precision of the parameters due to the addition of $ (\nu + \overline{\nu}) $ run from T2K and ICAL@INO.}
\label{fig:nova_deg}
\end{figure}

\vspace{-2mm}
In this section, we present the degeneracies present in both probability and $\chi^{2}  $ level and the role of the $ \overline{\nu} $ s to resolve these degeneracies. We also discuss the role of T2K and ICAL@INO. The first column of fig.(\ref{fig:nova_deg}) describes the degeneracies in the appearance channel.  The descriptions of the various bands  are given in the figure. We see here that the overlapping regions between  navy-blue and green bands show the degeneracy for the same values of $ \delta_{CP} $. Whereas, by drawing a horizontal line for a given probability one can identify various other degeneracies present at the probability level.
%%%%%%%%%%%%%%%%%%%%%%%%%%%%%%%%%%%%%%%%%%%%%%%%%%%%%%%%%%%%%%%%%%%%%%%%%
%/******************************************************
%\vspace{-0.3cm}
%%%%%%%%%%%%%%%%%%%%%%%%%%%%%%%%%%%%%%%%%%%%%%%%%%%%%%%%%%%
Second column shows the degeneracies at the $\chi^{2}  $ level. In this figure, the true point ($ 39^\circ,-90^\circ $) is marked with the black dot and the contours around it show the true solutions. Whereas, contours around $ \theta_{23} \sim 52^\circ$ show the degeneracies with wrong octant. In the third column we show the allowed region using NO$ \nu $A+T2K+ICAL@INO. In this case the degenerate solutions are removed by $ \overline{\nu} $-run and the allowed area is further constrained by T2K+ICAL@INO.
%$ (\nu + \overline{\nu}) $ run with .
 The detailed analysis for other sets of parameter values are presented in Ref\cite{Ghosh:2015ena}.
%$$$$$$$$$$$$$$$$$$$$$$$$$$$$$$$$$$$$$
%%%%%%%%%%%%%%%%%%%%%%%%%%%%%%%%%%%%%%%%%%%%%%%%%%%%%%%%%%%%%%%%%%%%%%%%%

\begin{figure}
\vspace{-0.5cm}
        \begin{tabular}{lr}
               \hspace*{0.3in}
\includegraphics[scale=0.7]{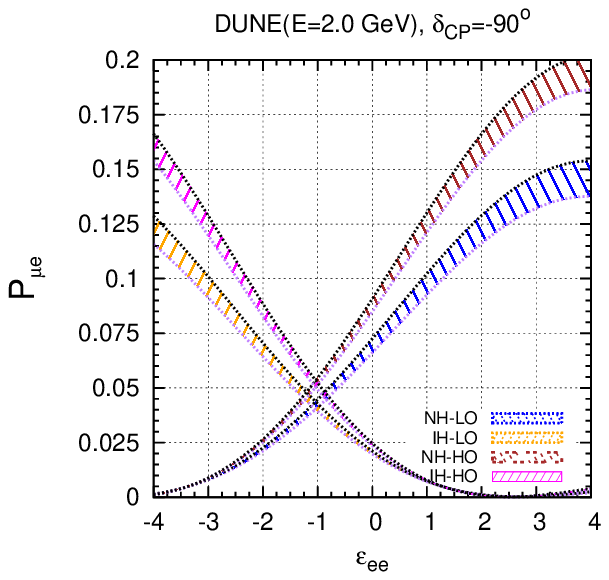}
\hspace*{-0.6in}
\includegraphics[scale=0.7]{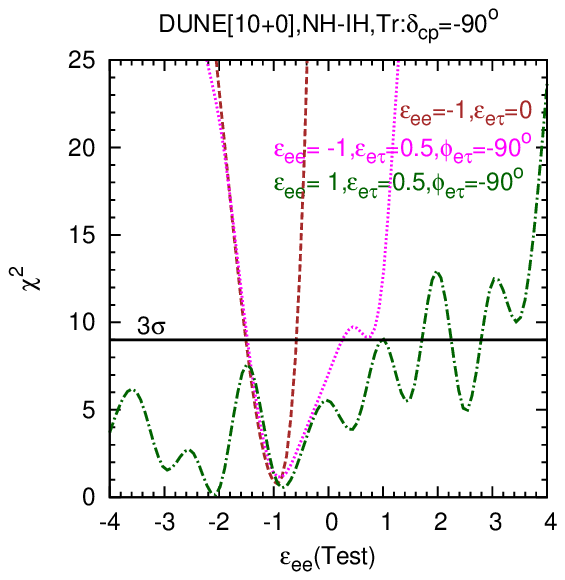}
        \end{tabular}
\vspace{-0.4cm}
\caption{\footnotesize Here, first column shows the appearance channel probability in presence of NSI parameter, $ \epsilon_{ee} $ for DUNE.  Whereas second column shows the hierarchy sensitivity in presence of NSI parameters $ \epsilon_{ee},~ \epsilon_{e \tau}  $.}
\label{fig:nsi_dune}
\end{figure}

In the fig.(\ref{fig:nsi_dune}), we describe the role of  NSI on the determination of hierarchy for DUNE. We focus on the effects of propagation NSI for which an extra contribution to the Lagrangian can come from  dimension-six four-fermion operators :
%%%%%%%%%%%%%%%%%%%%%%%%%%%%
\begin{equation}
  \label{eq:NSI}
 - \mathcal{L}^{NC}_\text{NSI} = 2\sqrt{2}G_F
   \epsilon^{fC}_{\alpha\beta}(\overline{\nu}_\alpha \gamma^{\rho} P_L \nu_\beta )
        ( \bar{f} \gamma_{\rho} P_C f ) + \text{H.c.}
\end{equation}
%%%%%%%%%%%%%%%%%%%%%%%%%%%%%%
where $\epsilon^{f C}_{\alpha\beta}$ are NSI parameters $\alpha, \beta = e, \mu, \tau$, $f = u,d,e$, $C$ denotes the chirality and $ G_{F} $ is the Fermi constant.
In  Ref\cite{Deepthi:2016erc}, we discussed the role  of the diagonal NSI parameter $ \epsilon_{ee} $.
%/******************************************************
%\vspace{-0.3cm}
%%%%%%%%%%%%%%%%%%%%%%%%%%%%%%%%%%%%%%%%%%%%%%%%%%%%%%%%%%%
In the first column of fig. \ref{fig:nsi_dune}, we present the $ P_{\mu e} $ vs $\epsilon_{ee}  $ (model-independent range) for fixed energy, $ \delta_{CP} $. The width of the bands are over octant for a given hierarchy as described in the figure. We mainly focus on a special point, $\epsilon_{ee}=-1  $ for which, the NSI effect gets nullified by the usual matter term. Hence, in absence of off-diagonal NSI parameters any LBL experiments will not be able to lift this degeneracy. In the second column, we discuss this degeneracy at the $ \chi^{2} $ level and also describe the role of off-diagonal NSI parameter $ \epsilon_{e \tau} $. Here, brown curve shows  the  degeneracy for $\epsilon_{ee}=-1  $ at $ \chi^{2} $ level and  the pink curve  shows that addition of $ \epsilon_{e \tau} $ is not able to lift the degeneracy once $\epsilon_{ee}=-1  $. Whereas, the green curve shows that if $\epsilon_{ee}\neq-1 $ then DUNE can have hierarchy sensitivity if certain ranges of $\epsilon_{ee}  $ are not allowed.
%and which will constraints parameter space of $\epsilon_{ee}  $.
%

In conclusion, we describe the (hierarchy$-$octant$-\delta_{CP} $) generalized degeneracy and their resolution using neutrino oscillation experiments. We also discuss the impact of NSI on the mass hierarchy determination in case of DUNE.
%, where we mainly focus on the role of digonal NSI. 

\end{document}